\documentclass[a4paper,12pt]{article}
\usepackage{epsfig}
\usepackage{graphicx}
\begin{document}
\title{Activity autocorrelation in financial markets. A comparative study between several models}
\author{Luigi Palatella\thanks{luigi.palatella@df.unipi.it}\\
INFM - Dipartimento di Fisica dell'Universit\`a di Pisa,\\ via Buonarroti 2, 56127 Pisa, Italy\\
\and
Josep Perell\'o\thanks{josep.perello@ub.edu}, Miquel Montero\thanks{miquel.montero@ub.edu}, and Jaume Masoliver\thanks{jaume.masoliver@ub.edu} \\
Departament de F\'{\i}sica Fonamental, Universitat de Barcelona,\\
Diagonal, 647, 08028-Barcelona, Spain
}

\maketitle

\begin{abstract}
We study the activity, {\it i.e.}, the number of transactions per unit time, of financial markets. Using the diffusion entropy technique we show that the autocorrelation of the activity is caused by the presence of peaks whose time distances are distributed following an asymptotic power law which ultimately recovers the Poissonian behavior. We discuss these results in comparison with ARCH models, stochastic volatility models and multi-agent models showing that ARCH and stochastic volatility models better describe the observed experimental evidences.
\end{abstract} 

\section{Overview}

As is well known, financial time series present a strongly inhomogeneous time behavior. This is specially true when one considers either the volatility or the activity, {\it i.e.}, the number of transactions per unit of time. Indeed if we look at the variance of the return in a time window of, say, one day, we will observe periods of relative constant and regular behavior followed by other periods of strong variation of the price. In the same way there are days with few transactions and others where the number of trades is considerably larger. This great variability in the volatility or in the activity is generally referred to as volatility clustering or intermittency of volatility and activity. In this work we refer to both quantities. We will thus perform measures on the activity and use two volatility models: (i) ARCH models \cite{engle} and (ii) stochastic volatility (SV) models~\cite{stochasticvolatilitybook,pre}, where the relationship between volatility and activity is set by the usual assumption of proportionality between them \cite{bouchaud1,bouchaud2}.

As is well known, the time interval or distance between two consecutive transactions $\tau$ is a random variable described by a probability density function (pdf) $\psi(\tau)$ which in many cases presents an asymptotic power law of the form\cite{masoliver1,masoliver2}
\begin{equation}
\psi(\tau) \sim 1/\tau^{\beta}.
\label{powerlawpsi}
\end{equation}
However $\psi(\tau)$ does not tell anything about the independence of consecutive $\tau$'s. We note that if consecutive $\tau$'s are independent a power law tail in $\psi(\tau)$ can explain an inhomogeneous behavior in the number of events per unit of time, where a possible measure of this inhomogeneity is the distribution of the number of trades in a fixed period of time $t$. As Feller proved many years ago \cite{feller}, if the time interval $\tau$ between some particular events, which we will call markers, in a time series is distributed according to a given density $\psi(\tau)$ and the independence condition $\langle \tau_i \tau_j \rangle = \langle \tau_i \rangle \langle
\tau_j \rangle $ for $i \neq j$ holds, then the probability distribution to observe a fixed number of these markers $y$ in a given time interval, $p(y,t)$, follows a scaling law of the form
\begin{equation}
p(y,t)=\frac{1}{t^{\delta}} F \left ( \frac{y}{t^{\delta}}\right),
\label{scalinglaw}
\end{equation}
where $\delta$ is some positive exponent and $F(x)$ is a positive and integrable function. With the help of a recently developed technique for the analysis of time series called Diffusion Entropy (DE) \cite{giacomo}, we will see that the scaling observed in the distribution of the number of transactions in a time interval does not correspond to Feller's analytical prescription obtained with the density $\psi(\tau)$ estimated from data.

We are thus forced to make the additional hypothesis that consecutive $\tau$'s are not independent. We will also assume that this correlation is due to the presence of peaks (or clusters) in the mean activity followed by periods of relative calm. Therefore the $ \tau_i$'s are positively correlated because during a peak of activity they are shorter than the mean value while away from a peak they are greater than the mean. Indeed, in such a case $\langle(\tau_i-\langle\tau_i\rangle)(\tau_j-\langle\tau_j\rangle)\rangle\geq 0$ which implies a positive correlation: $\langle\tau_i\tau_j\rangle\geq\langle\tau_i\rangle\langle\tau_j\rangle$.

Let $\tau_c$ be the random time distance between two consecutive peaks and denote by $\phi(\tau_c)$ its probability density function. Similarly to the distribution of the distance between two consecutive transactions $\psi(\tau)$, we will also assume that $\phi(\tau_c)$ obeys an asymptotic power law:
\begin{equation}
\phi(\tau_c) \sim 1/\tau_c^{\mu}.
\label{powerlaw}
\end{equation}
In this scheme the results of the DE technique can be described directly in
terms of the time distance and the magnitude of the peaks of activity, the latter described by a pdf $h(x)$. We will see, like in ref. \cite{terremoti}, that the distribution of the size of the cluster given by $h(x)$ does not play an
important role, because the time distance distribution $\phi(\tau_c)$ is characterized
by a more anomalous exponent than $h(x)$. Consequently we will interpret the
results of DE as a consequence of a non-Poissonian distribution of the distance between peaks of activity. 

There are in the literature several approaches that try to explain the autocorrelation of activity and volatility. One recent model \cite{bouchaud1,bouchaud2,lux} is based on the hypothesis that the intermittency of activity is caused by a subordination to
a random walk, like in the case of the so called on-off intermittency
\cite{on-off}. As clearly described in \cite{bouchaud2}, this procedure should give for the distribution of distances between clusters a scaling law of the form
\begin{equation}\label{eqbouchaud}
\phi(\tau_c) \simeq \frac{1}{\tau_c^{3/2}} f \left (
\frac{\tau_c}{\lambda} \right )
\end{equation}
where $f(t)$ is a cutoff function ensuring the existence of the first moment of $\phi(\tau_c)$ and $\lambda$ is the time scale at which
this cutoff takes place. One simple choice for $f(t)$ is given by the exponential 
$f(t)=e^{-t}$ which allows that $\phi(\tau_c)$ presents an asymptotic Poissonian behavior.

As we have already mentioned, other possible approaches to the problem of activity correlation are provided by ARCH models or SV models. We will show that both, ARCH and SV, models lead to a correlation in the volatility which more likely resembles to a power law tail exponent observed in a variety of financial markets. We will also show that a particular ARCH model, the TARCH model presented in \cite{englepatton}, and the SV model presented in \cite{pre,masoliver3} both result in the same scaling law than that observed with the DE technique. Finally, and due to the absence of intra-day disturbances in the time series obtained by ARCH and SV models, we are also able to evaluate numerically the waiting time distribution $\phi(\tau_c)$ of the distance between peaks and this distribution is compatible with the empirical evidence given by a power-law behavior for a (long) transient period followed by a Poissonian (exponential) behavior. We incidentally note that this asymptotic exponential behavior indicates that very far clusters do not influence each other.

The paper is organized as follows. We start with a brief review on the DE technique and a simple analytical proof that the DE results are determined by the most
anomalous power law tail between $\phi(\tau_c)$ and $h(x)$. After
that we show the results obtained by means of the DE technique on tick by tick data of a Foreign Exchange (FX) market. We also perform a filtering procedure on data in order to prove that the observed scaling is due to the anomaly in the waiting time pdf $\phi(\tau_c)$ and not in the cluster size pdf $h(x)$. Finally, we briefly describe the ARCH and the SV models and the results of the DE and the waiting time distribution on the time series constructed using these models.

\section{Diffusion entropy analysis}

\begin{figure}
\begin{center}
\includegraphics[width=0.9\textwidth]{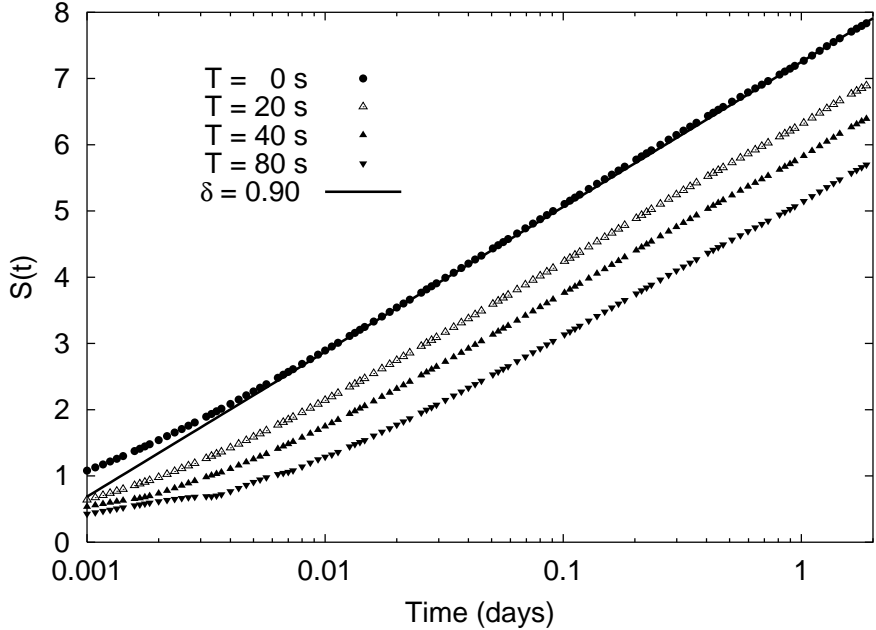}
\caption{\label{fig1b}Results of the DE analysis for US dollar - Deutsche mark
futures market for different values of the time-threshold $T$.}
\end{center}
\end{figure}

 The diffusion entropy technique is basically an algorithm designed to detect 
memory in time series \cite{giacomo}. DE is specially suitable for intermittent
signals, {\it i.e.,} for time series where bursts of activity are separated by periods of quiescent and regular behavior. The technique has been designed to study the time distribution of some markers (or events) along the time series and thus discover 
whether these events satisfy the independence condition $\langle \tau_i \tau_j \rangle = \langle \tau_i \rangle \langle
\tau_j \rangle $ ($i \neq j$) where $\tau_i$ is the time interval between the marker labeled $i-1$ and the next one $i$ \cite{terremoti,cuore1,cuore1b,cuore2}. As marker we use here a very simple definition: {\it each trade in the time series is a marker}. In order to apply the DE technique we need to construct a new series $\xi_i$ which is a function of a coarse grained time $i \cdot \Delta t$ (in our case $\Delta t= 1$ s) and where $\xi_i$ is precisely the number of transactions that occurred in the previous second. We next define a new random process through the following moving counting on $\xi_i$
\begin{equation}
y_l(t)=\sum \limits_{i=l-t/\Delta t}^{l} \xi_i.
\label{y(t)}
\end{equation}
Note that $y_l(t)$ is precisely the number of markers ({\it i.e.,} trades) in an interval of length $t$ starting at position $l$. If we vary the value of $l$ along the interval $[0,N-t/\Delta t]$, where $N$ is the total length of the sequence, we can obtain the probability density function, $p(y,t)$, of this random process. It has been shown in \cite{giacomo} that for the zero-mean process $y\rightarrow y -\langle y(t)\rangle$ and assuming that $\xi_i$ is a renewal process, then $p(y,t)$ obeys the scaling law given by Eq. (\ref{scalinglaw}). Hence, the entropy of this random process reads \cite{giacomo}
\begin{equation}
S(t)=-\int\limits_{-\infty}^{\infty} p(y,t) \ln \left[ p(y,t)
\right ] {\rm d}y = A +\delta \ln(t).
\label{S(t)}
\end{equation}
From the slope of $S(t)$, in a logarithmic scale, we get an estimation of
the scaling parameter $\delta$. In Fig. \ref{fig1b} we show the results obtained with DE on tick by tick data of the US dollar- Deutsche mark futures market from 1993 to 1997 with a total of $1.3 \times 10^6$ data points (solid circles) \cite{note1}. The fit gives $\delta= 0.90$. As is shown in \cite{giacomo} if condition $\langle\tau_i\tau_j\rangle=\langle\tau_i\rangle\langle\tau_j\rangle$ holds then there exists a relation between the scaling exponent $\delta$ and the exponent $\beta$ of the power law tail of the time distribution of markers (in our case, trades) $\psi(\tau)$ 
(see Eq. (\ref{powerlawpsi})). This relation reads
\begin{equation}
\delta= \cases{1/(\beta-1), &if \ $2<\beta<3$;\cr
 0.5, &if \ $\beta>3$.}
\label{giacomoeq}
\end{equation}
We have shown elsewhere that for the FX market under consideration the power law tail exponent of the waiting time distribution between trades is near $3.5$ \cite{masoliver1,masoliver2}. According to Eq. (\ref{giacomoeq}) this would lead to $\delta=0.5$ ---after a transient comparable with the mean trade distance of $\langle \tau \rangle = 23.6$ s--- in disagreement with the value $\delta= 0.90$ obtained by DE. 

\begin{figure}
\begin{center}
\includegraphics[width=0.9\textwidth]{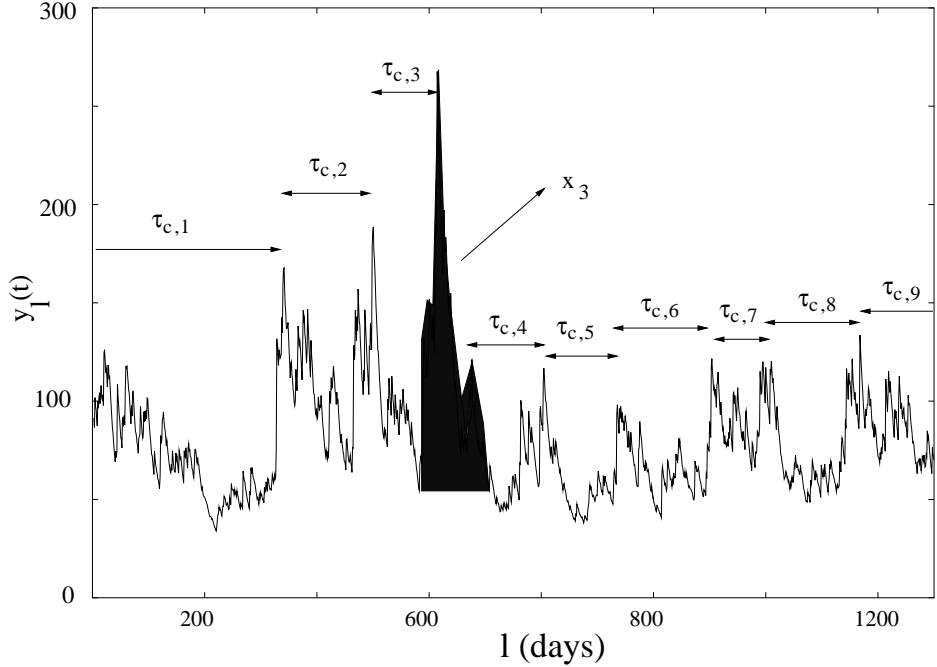}
\caption{\label{schema} Scheme of the model for the anomalous
scaling for activity and volatility.}
\end{center}
\end{figure}

We now present a picture which takes into account peaks of activity separated by periods with a low number of transactions. We see in Fig. \ref{schema} a schematic representation of this picture in which, for instance, the intensity $x_3$ of the third peak is represented by a black spot and this corresponds to the total number of transactions attributed to this peak. 

We suppose that the time intervals between peaks, $\tau_{c,i}$, are distributed according to a pdf $\phi(\tau_c)$ which asymptotically behaves as in Eq. (\ref{powerlaw}), {\it i.e.}, $\phi(\tau_c)\sim 1/\tau_c^\mu$. We also assume that the intensity of a given peak, $x_i$, defined as the total number of transactions in the peak, has a distribution given by a pdf $h(x)$. In the context of earthquakes $h(x)$ is generally referred to as the Pareto law of the size of the earthquake clusters, since asymptotically $h(x)\sim 1/x^{\alpha+1}$ as in the usual Pareto distribution \cite{terremoti}. 

We will now present a proof that the DE technique perceives the most anomalous ({\it i.e.}, the smallest) of the two exponents $\mu$ and $\alpha+1$. Indeed, let $\rho(x,t)$ be the joint pdf for the waiting time $\tau_c$ between clusters and their intensity $x$. We denote by 
$$
\widehat{\rho}(\omega,s)=\int_{-\infty}^{\infty}dx e^{i\omega x}\int_0^\infty
d\tau_ce^{-s\tau_c}\rho(x,\tau_c)
$$
its Fourier-Laplace transform. Observe that in terms of $\widehat{\rho}(\omega,s)$ the Laplace transform of the waiting time distribution $\widehat{\phi}(s)$ and the Fourier transform of the size distribution $\tilde{h}(\omega)$ are given by
\begin{equation}
\widehat{\phi}(s)=\widehat{\rho}(\omega=0,s)\qquad\mbox{and}\qquad
\tilde{h}(\omega)=\widehat{\rho}(\omega,s=0).
\label{marginals}
\end{equation}
We assume that the time duration of a peak is negligible with respect to the mean time distance between peaks \cite{note3}. In such a case we can use the Continuous Time Random Walk (CTRW) formalism to calculate the probability density function, $p(y,t)$, that the number of trades at time $t$ is given by $y$. Thus, in terms of the joint distribution $\widehat{\rho}(\omega,s)$ the Fourier-Laplace transform of $p(y,t)$ is given by 
\cite{masoliver1,masoliver2}
\begin{equation}\label{ctrw}
\widehat{p}(\omega,s) = \frac{1-\widehat{\phi}(s)}{s}
\frac{1}{1-\widehat{\rho}(\omega,s)}.
\end{equation}

We can easily see that as $s\rightarrow 0$ ({\it i.e.}, $t\rightarrow\infty$) 
$[1-\widehat{\phi}(s)]/s\sim\langle\tau_c\rangle$, where $\langle\tau_c\rangle$ is the mean waiting time. Hence
\begin{equation}
\widehat{p}(\omega,s) \simeq\frac{\langle\tau_c\rangle}{1-\widehat{\rho}(\omega,s)},
\qquad(s\rightarrow 0).
\label{1stapprox}
\end{equation}
Note that, as $s\rightarrow 0$ we have (see Eq. (\ref{powerlaw}))
\begin{equation}
\widehat{\phi}(s) \simeq 1-\langle\tau_c\rangle s + c_0 s^{\mu-1},\qquad(2<\mu<3).
\label{phi(s)}
\end{equation}
Moreover, as $\omega\rightarrow 0$ and consistently with the Pareto law according to which $h(x)$ decays as $1/x^{\alpha+1}$, we have
\begin{equation}
\tilde{h}(\omega)\simeq 1+i\langle x\rangle\omega+b_0 \omega^{\alpha},\qquad(1<\alpha<2), 
\label{h(k)}
\end{equation}
where $\langle x\rangle$ is the average peak intensity. Taking into account Eqs. (\ref{phi(s)})-(\ref{h(k)}) we see that as $s\rightarrow 0$ and $\omega\rightarrow 0$ the joint distribution $\widehat{\rho}(\omega,s)$ can be written as 
\begin{equation}
\widehat{\rho}(\omega,s)\simeq 1-\langle \tau_c\rangle s+i\langle x\rangle\omega-
i\langle x\tau_c\rangle\omega s+b(s)\omega^{\alpha}+c(\omega)s^{\mu-1},
\label{rhoapprox}
\end{equation}
where $b(s)$ and $c(\omega)$ are such that $b(0)=b_0$ and $c(0)=c_0$. We recall that the DE technique measures the scaling in a 
``moving reference frame" where the average activity is zero, $\langle y(t) \rangle =0$ for all $t\geq 0$. In order to obtain the pdf for $y(t)$ in such reference frame we perform in Eq. (\ref{rhoapprox}) the following substitution
\begin{equation}\label{trasla}
s \longrightarrow s+i\omega\frac{\langle x\rangle}{\langle\tau_c\rangle},
\end{equation}
and after applying the diffusive limit $\langle\tau_c\rangle|s|\ll\langle x\rangle|\omega|$ we get, to the lowest order,
\begin{equation}
\widehat{\rho}(\omega,s)\simeq 1-\langle\tau_c\rangle s+b_0\omega^{\alpha}+
c_0(i\langle x\rangle/\langle\tau_c\rangle)^{\mu-1}\omega^{\mu-1}.
\label{2ndapprox}
\end{equation}
Substituting Eq. (\ref{2ndapprox}) into Eq. (\ref{1stapprox}) finally yields
\begin{equation}\label{grande}
\widehat{p}(\omega,s)\simeq\frac{\langle\tau_c\rangle}
{s\langle\tau_c\rangle-b\omega^{\alpha}-
c\left({i\langle x\rangle}/{\langle\tau_c\rangle}\right)^{\mu-1}\omega^{\mu-1}}.
\end{equation}
This equation shows that the smallest exponent between $\alpha$ and $\mu-1$ determines the asymptotic scaling of $p(y,t)$ according to the exponent 
\begin{equation}
\delta=\cases{1/(\mu-1), &if \ $\mu-1>\alpha$;\cr
 1/\alpha , &if \ $\mu-1<\alpha$.}
\label{mostanomalous}
\end{equation}
Therefore, the scaling perceived by DE is determined by the most anomalous exponent of the scaling between the size of the clusters of activity and the distribution of their time distances. Note that the case $\delta=1/(\mu-1)$ agrees with that of Eq. (\ref{giacomoeq}). We also observe that we have proven this fundamental result for the most general case in which there is no assumption on the possible correlation, or independence, among intensities and waiting times.

Having this in mind, we return to the problem of understanding the scaling exponent $\delta=0.90$ appearing in the US dollar - Deutsche mark futures market. To what effect is due this scaling? In other words, is the exponent $\delta$ determined by the time distance between clusters or by their size? In order to solve this question we impose a cutoff in the size of the peaks of activity by eliminating those transactions whose time distance from the previous one is below certain threshold $T$ (note that this actually reduces cluster sizes because the number of transactions counted is now smaller). If after this cutoff procedure the scaling remains invariant then $\delta$ would be 
determined by the time distances and not by the size of the clusters. In Fig. \ref{fig1b} the DE results are shown for different values of the time-threshold $T$ ranging from $0$ to
$80$ s. We see there that the slope is practically unchanged which confirms the assumption that the exponent $\delta=0.90$ is solely determined by the anomaly in the time distances between the clusters and not by any anomaly of their size.

\section{ARCH, Stochastic volatility and on-off intermittency models}

At the end of the last section, we have indirectly shown that the anomalous scaling $\delta=0.90$ observed in data is not caused by fat tails in the peak intensity distribution $h(x)$ but by the anomalous scaling in the waiting time distribution between peaks. Another more direct way to prove this would have been to single out the peaks on real data and look for their time distribution. Unfortunately it is very difficult, on real data, to define a peak of activity and compute the waiting time distribution between them. This is because there are peaks of activity that appear at fixed times (we will call these ``deterministic peaks") at the daily opening and closing sessions, at the opening during the day of other markets and even weekly at the opening of each monday \cite{FFT}. These deterministic peaks do not contribute to the increase of entropy. However, they do affect any estimation of the waiting time distribution making it very difficult to get a reliable estimate of it. A possible way out from this situation would be to generate an artificial time series simulating the real market evolution. In this artificial series, the activity would be replaced by volatility following the accepted correspondence between them~\cite{bouchaud1,bouchaud2} and we would check there all the scaling phenomena reported up till now. 

\begin{figure}[h]
\begin{center}
\includegraphics[width=0.9\textwidth]{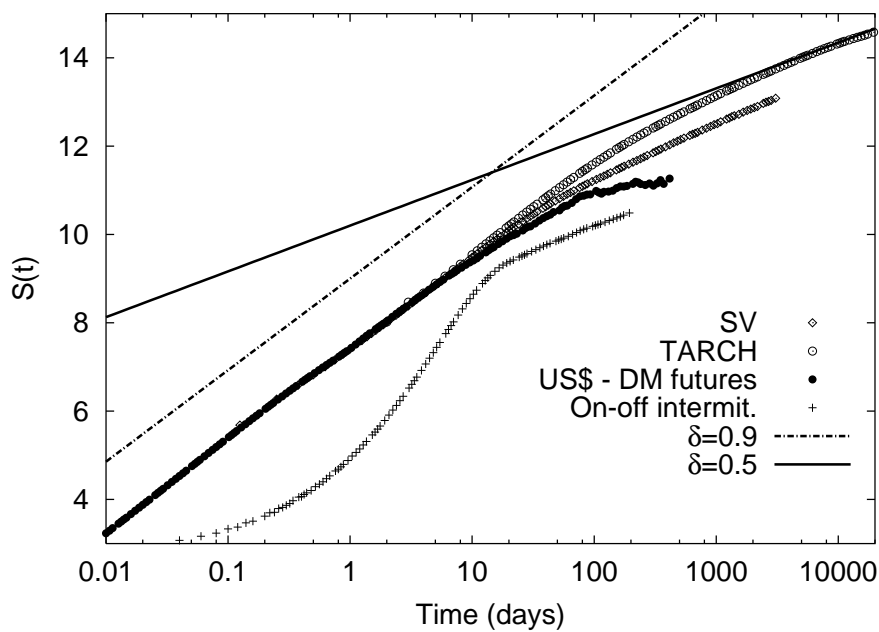}
\caption{\label{fig1} Results of the DE analysis for US dollar - Deutsche mark futures with $T=0$ (solid circles), the TARCH model (empty circles), the SV model (diamonds), and finally for the on-off intermittency model as given by Eq. (\ref{eqbouchaud}) (crosses).}
\end{center}
\end{figure}

We will follow this procedure and choose two well accepted models for reconstructing market activity without deterministic peaks: (i) the TARCH model \cite{englepatton}, and (ii) the stochastic volatility model presented in \cite{pm}. We will see that both models give the same results than those of the DE technique. We finally discuss the prescription given in Eq.(\ref{eqbouchaud}) based on multi-agent models to see whether it agrees with our results or not. 

\subsection{The TARCH model}

This particular ARCH model, called TARCH by its authors \cite{englepatton}, is given by 
\begin{eqnarray}\label{TARCH}
 \sigma^2_{t} & = & k + \alpha R_{t-1}^2 + \beta \sigma^2_{t-1} +
 \Theta(- R_{t-1}) \gamma R_{t-1}^2 \\
R_{t} & = & \sigma_t \eta_{t} \nonumber
\end{eqnarray}
where $\sigma_t$ is the volatility, $R_t$ is the one day return
calculated at time $t$, $\Theta(x)$ is the Heavyside step function
and $\eta_t$ is Gaussian noise with zero mean and unit variance. The other parameters, estimated from daily data of the Dow Jones Industrial Index from 1988 to 2000 and obtained in \cite{engle}, are: $k = 0.0184 $, $\alpha = 0.0151 $, $\gamma =
0.0654$, $\beta = 0.9282$ \cite{note2}. Using Eq. (\ref{TARCH}) we generate a time series for $\sigma_t$. We then perform the DE analysis on this series (with a time step
of 1 day) by supposing that the number of trades in the i-th day is proportional to $\sigma_i$. The results are shown in Fig.~\ref{fig1} (empty circles) compared with the results on real data for $T=0$ (solid circles). We clearly see that the TARCH model predicts for $t<t_P$ a scaling exponent $\delta=0.90$ which agrees with actual data. For $t$
greater than a Poissonian time $t_P \approx 100 $ days the model yields $\delta=0.5$. It is worth noticing that the change in the slope of real data is likely due to the lack of statistics. Moreover, we do not have enough data points to determine whether the change of
slope takes place at the same time scale than in the TARCH model. Nevertheless, ARCH-type models (similar results were obtained with $\gamma=0$ in Eq.(\ref{TARCH})) seem to take into account the correct structure of the intermittency of financial series.

\begin{figure}[h]
\begin{center}
\includegraphics[width=0.9\textwidth]{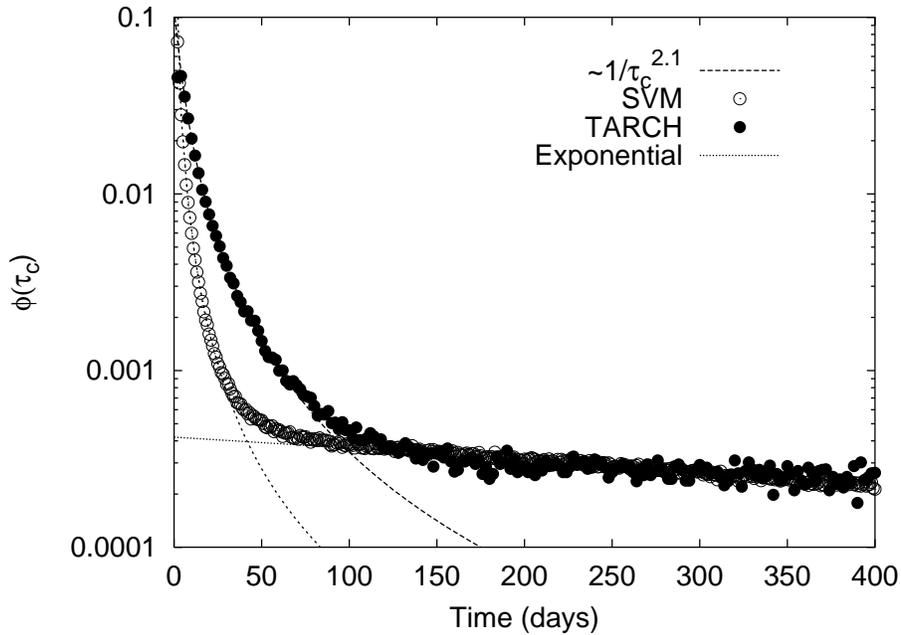}
\caption{\label{fig2} Waiting time distribution for the distance between clusters $\tau_c$ (in logarithmic scale) for the TARCH model (solid circles) and the SV model (empty circles). A distinct asymptotic behavior is clearly present: a power law tail with exponent $\mu \simeq 2.1$ and an exponential decay.}
\end{center}
\end{figure}

From the series generated using Eq. (\ref{TARCH}) we can also evaluate the waiting time distribution between peaks because now we do not have deterministic peaks and other periodic effects that were present in actual data. The result is shown in Fig.~\ref{fig2} and as we see there that for $\tau_c<t_P$ a good fit is provided by the following power law: 
$$
\phi(\tau_c) \simeq \frac{1}{(1+\nu\tau_c)^{\mu}},
$$
where $\nu=0.1\mbox{ day }^{-1}$ and $\mu=2.1$. For $\tau_c>t_P$ a clear exponential (Poisson) behavior is present. This result has
a simple physical explanation: if there is a first cluster at time $t$ the probability to observe another one just after the first is high while very distant clusters are
practically independent which explains the asymptotic Poisson observed behavior in 
Fig.\ref{fig2}. 

\subsection{An stochastic volatility model}

There exists another way of modelling volatility clustering. The so-called stochastic volatility models \cite{stochasticvolatilitybook,masoliver3,pre} are an alternative choice to
ARCH models and they are considered to be the most natural extension to the classic geometric Brownian motion for the price dynamics in continuous-time finance. Let us start with the zero-mean return $X(t)$ (i.e., the log-price without drift) and whose dynamics
is given by the following stochastic differential equation 
\begin{equation}
\dot{X}(t)=\sigma \xi_1(t), 
\label{xeq}
\end{equation}
where this equation has to be understood in the It\^o sense and $\sigma$ is the volatility and $\xi_1dt$ is the Wiener process, {\it i.e.}, $\xi_1(t)$ is Gaussian white noise with zero mean and correlation function given by 
\begin{equation}
\langle\xi_1(t)\xi_1(t')\rangle=\delta(t-t'). 
\label{xi}
\end{equation}
All SV models assume that the volatility $\sigma$ is itself a random process. There
are several ways to describe the dynamics of the volatility 
\cite{stochasticvolatilitybook}. One of the simplest models, which still contains almost all the basic ingredients prescribed by real markets, is given by the Ornstein-Uhlenbeck 
(OU) process~\cite{masoliver3}
\begin{equation}
\dot{\sigma}(t)=-a(\sigma-m)+k\xi_2(t). \label{sigma}
\end{equation}
One key property of this model is that it exhibits a stationary solution thanks to the existing reverting force ---quantified by $a$--- to a certain average $m$, the so-called ``normal level of volatility". The stationary solution is a Gaussian distribution and the resulting distribution for the return has fat tails \cite{masoliver3}. In addition, stylized facts such as the negative skewness and the leverage correlation 
\cite{masoliver3,pre} require that the changes of the volatility be negatively correlated with the random source of return changes. In other words, the driving noises appearing in Eqs. (\ref{xeq}) and (\ref{sigma}) are anti correlated, that is:
$$
\langle\xi_1(t) \xi_2(t')\rangle= \rho\delta(t-t')
$$
where $-1\leq\rho\leq 0$. For the OU SV as given in Eq. (\ref{sigma}) the characteristic exponential time decay of leverage correlation is given by $1/a$ which is typically of the order of few trading days (see below).

Although the OU model has some disagreements with observations~\cite{pm}, it is complex enough to catch all the statistical properties that we are studying here. We therefore simulate the SV model with the parameters estimated from daily data of the Dow
Jones Index from 1900 to 1999. Thus, the reverting force is equal to $a=0.05\mbox{ days }^{-1}$, the noise amplitude is $k=0.0014\mbox{ days }^{-1}$, the normal level of volatility reads $m=0.011\mbox{ days }^{-1/2}$ and the correlation coefficient is $\rho=-0.5$~\cite{masoliver3}. The results of the DE analysis are reported in Fig. \ref{fig1} (diamonds) while the waiting time distribution between clusters is shown in Fig. \ref{fig2} (empty circles). In this case, and analogously to the TARCH model, we also observe a power law behavior followed by an exponential decay. The only difference is the value of the Poissonian time $t_P$ which for this model is near to 40 days while for the TARCH model is approximately 100 days. We have checked numerically that this difference is due to the fact that the parameters defining each model are estimated from the Dow Jones index {\em over a different period of time}, much larger for the SV model than for TARCH model. In any case, we cannot discard any of these approaches on the basis of the empirical results.
%In any case, and as we have mentioned, it has been impossible to experimentally confront which measure of the Poissonian time $t_P$ is more accurate due to the lack of data eventhough the measure given by the SV model should be more accurate since parameter estimation is more reliable for SV than for TARCH. 

\subsection{On-off intermittency models}

The intermittent model of the activity is also predicted and studied by several multi-agent or minority game models \cite{bouchaud1,bouchaud2,helbing}. These models can be connected to on-off intermittency and they generally imply that the persistency of activity is subordinated to a random walk which indicates that the waiting time distribution has the form given in Eq.(\ref{eqbouchaud}). As suggested in \cite{bouchaud1,bouchaud2} we also obtain the so-called variogram of the data, although the results denote that, if a
$3/2$ tail is present like in Eq.(\ref{eqbouchaud}), causing a $\sqrt{t}$ behavior in the variogram, the tail lasts less than 5 days probably because the FX market is more liquid than the ones considered in \cite{bouchaud1,bouchaud2}. Furthermore the DE analysis performed on a series generated according to Eq. (\ref{eqbouchaud}) also leads to a transient followed by the exponent $\delta=0.5$ of the asymptotic behavior whereas, as is clearly seen in Fig. \ref{fig1}, the transient never exhibits the exponent $\delta=0.90$ beyond 1 day but presents a constantly increasing exponent which is most of the time greater than 1.

\section{Conclusions}

We have performed the DE analysis on the activity of the tick by tick time series US dollar - Deutsche mark futures from 1993 to 1997. The results clearly show the presence of an anomalous scaling, for the probability distribution of the activity $p(y,t)$, near the exponent $\delta=0.90$. We have also implemented the same analysis on the volatility obtained with the TARCH model and with the OU stochastic volatility model. We find in both cases an excellent agreement between the scaling measured either on the actual data and on the constructed series.

We compare the results with the scheme of the subordination of the volatility to a random walk leading to Eq.~(\ref{eqbouchaud}) observing that a power law exponent $\mu\approx 2$ for the tail of the distribution of the distances between peaks $\phi(\tau_c)$ of volatility is more plausible. We believe that the main reason why the TARCH and the SV models give better results is that in on-off intermittency models the occurrence of a peak can be considered as subordinated to a random walk but the weak restoring force (which has to be included in the model in order to describe mean reversion) not only causes the final stationarity and the Poisson tail of Fig.~\ref{fig2} but also affects the process of
regression to equilibrium modifying in a fundamental way (from $\mu=3/2$ to $\mu\simeq 2$) the transient behavior of $\phi(\tau_c)$.

%\ack
\section*{Acknowledgements}
This work has been supported in part by Direcci\'on General de Proyectos de Investigaci\'on under contract No. BFM2003-04574, and by Generalitat de Catalunya under contract No. 2001SGR-00061.


\begin{thebibliography}{99}
\bibitem{note1} Data record is made up of 6 hours and 40 minutes of trading activity each day. The opening of a given day has been attached to the closure of the previous day thus neglecting any possible over-night dynamics.
\bibitem{note3} We have checked numerically that this hypothesis should not be considered as a real constraint. Even in the case that the typical time duration of a peak is comparable with the distance between peaks, the condition required in DE technique for obtaining the correct asymptotic scaling is that $t\gg\langle\tau_c\rangle$.
\bibitem{FFT} This periodic behavior can be easily singled out by a fast Fourier transform of the FX data.
\bibitem{note2} In the original model there is a mean drift in the return corresponding to the mean growth of the market. Our data correspond to a FX market where this effect is not present, Therefore, there is no mean drift in our analysis.
\end{thebibliography}
\end{document}